\documentclass[letter]{aa} 
\usepackage{natbib}
\usepackage{graphicx}
\usepackage{txfonts}
\usepackage{epsfig}
\usepackage{times}
\usepackage{rotating}
\usepackage{float}
\usepackage[caption = false]{subfig}
\usepackage{setspace}
\usepackage{lscape}
\usepackage{epstopdf}
\usepackage{tabularx}
\usepackage{caption}

\DeclareCaptionFormat{cont}{#1 (cont.)#2#3\par}

\begin{document}

\title{Fossil group origins}

\subtitle{XIII. A paradigm shift: fossil groups as isolated structures rather than relics of the ancient Universe.}

\authorrunning{S. Zarattini et al.}

\titlerunning{The peculiar position of FGs within the cosmic web}

\author{S. Zarattini\inst{1,2,3,4}, J. A. L. Aguerri\inst{1,2}, P. Tarr\'io\inst{5}, and E. M. Corsini\inst{3,4}}

\institute{Instituto de Astrof\'isica de Canarias, calle Vía L\'actea s/n, E-38205 La Laguna, Tenerife, Spain
\email{szarattini@iac.es}
\and Departamento de Astrof\'isica, Universidad de La Laguna, Avenida Astrof\'isico Francisco S\'anchez s/n, E-38206 La Laguna, Spain
\and Dipartimento di Fisica e Astronomia ``G. Galilei'', Universit\`a di Padova, vicolo dell'Osservatorio 3, I-35122 Padova, Italy
\and INAF-Osservatorio Astronomico di Padova, vicolo dell'Osservatorio 5, I-35122 Padova, Italy
\and Observatorio Astron\'omico Nacional, IGN, Calle Alfonso XII 3, E-28014 Madrid, Spain}

\date{\today}

\abstract{}
{In this work we study the large-scale structure around a sample of non-fossil systems and compare the results with earlier findings for a sample of genuine fossil systems selected using their magnitude gap.}
{We compute the distance from each system to the closest filament and intersection as obtained from a catalogue of galaxies in the redshift range $0.05 \le z \le 0.7$. We then estimate the average distances and distributions of cumulative distances to filaments and intersections for different bins of magnitude gap.}
{We find that the average distance to filaments is $(3.0\pm 0.8)$ $R_{200}$ for fossil systems, whereas it is $(1.1\pm 0.1)\,R_{200}$ for non-fossil systems. Similarly, the average distance to intersections is larger in fossil than in non-fossil systems, with values of $(16.3\pm 3.2)$ and $(8.9\pm 1.1) \,R_{200}$, respectively. Moreover, the cumulative distributions of distances to intersections are  statistically different between fossil and non-fossil systems.}
{Fossil systems selected using the magnitude gap appear to be, on average, more isolated from the cosmic web than non-fossil systems. No dependence is found on the magnitude gap (i.e. non-fossil systems behave in a similar manner independently of their magnitude gap and only fossils are found at larger average distances from the cosmic web). This result supports a formation scenario for fossil systems in which the lack of infalling galaxies from the cosmic web, due to their peculiar position, favours the building of the magnitude gap via the merging of all the  massive satellites with the central galaxy. Comparison with numerical simulations suggests that fossil systems selected using the magnitude gap are not old fossils of the ancient Universe, but systems located in regions of the cosmic web not influenced by the presence of intersections.}


\maketitle

\section{Introduction}
\label{sec:intro}
\citet{Ponman1994} proposed for the first time the existence of fossil groups (FGs) to explain the discovery of an apparently isolated giant elliptical galaxy, which was surrounded by an extended halo typical of a galaxy group. They supposed that this was the fossil relic of an old group of galaxies, which had enough time to merge all its bright galaxies within the brightest central one (BCG). FGs should be older that regular groups in order to merge all the bright galaxies and create giant isolated BCGs. They were thus proposed as fossil relics of the ancient Universe.

\begin{figure*}
    \centering
    \includegraphics[trim=80 10 200 60,width=0.8\textwidth]{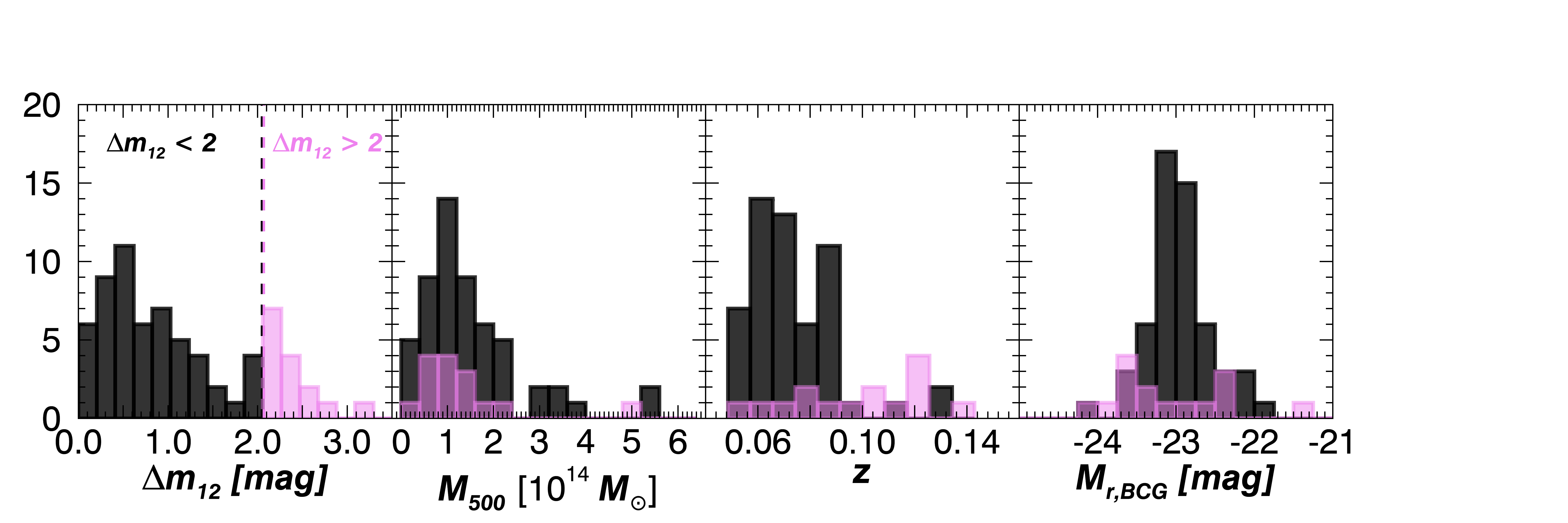}
    \caption{Properties of the sample systems. From left to right: magnitude gap distribution, total mass distribution,  redshift distribution, and distribution of the absolute magnitude of the BCGs. The FGs from \citet{Zarattini2022} are shown in violet, while the non-FGs are shown in black.}
    \label{fig:sample}
\end{figure*}

The first observational definition of FGs was given by \citet{Jones2003}. They define an FG as a system with a large magnitude gap between the two brightest members ($\Delta m_{12} > 2$ mag in the $r$ band) within half the virial radius of the group. 
The number of FGs slowly but steadily grew along the years. \citet{Vikhlinin1999} proposed four candidates defined as ``over luminous red galaxies'', \citet{Jones2003} presented a sample of five FGs, then \citet{Santos2007} listed the first large sample of 34 FG candidates, of which 15 were confirmed as genuine FGs by \citet{Zarattini2014}. \citet{Miller2012} and \citet{Adami2018} identified 12 and 18 FGs respectively and, more recently, \citet{Adami2020} has proposed a new probabilistic method to find FGs, favouring the statistical studies of these systems. Finally, in their review \citet{Aguerri2021} collected a list of 36 genuine FGs from the literature. These are the systems for which $\Delta m_{12}$ was spectroscopically measured and the virial radius was properly computed.

However, several observational evidences showed that FGs are not older than non-FGs. For example, the fraction of galaxy substructure in FGs is similar to that of non-FGs \citep[][]{Zarattini2016} and the stellar age of the central galaxies in FGs is similar or even younger than those in non-FGs \citep[][]{LaBarbera2012,Eigenthaler2013,Corsini2018}. In a very recent work, \citet{Chu2023} studied a sample of central galaxies in FGs and non-FGs candidates, finding that BCGs in FGs should have evolved in a similar way as regular BCGs. This result was also confirmed by their analysis of the stellar populations. In fact, they claimed that the stellar populations can not be distinguished between BCGs in FG and non-FG candidates. In addition, cosmological simulations pointed out that the galaxy aggregations selected as FGs by using the magnitude gap did not form at earlier epochs than non-FGs \citep[][]{Kanagusuku2016,Kundert2017}. On the other hand, \citet{Raouf2014} suggested that old FGs could be those with large magnitude gaps and small central galaxies. 

If FGs are not older than non-FGs, another mechanism is required to speed up the merging rate of massive satellites in these systems. \citet{Sommer-Larsen2006} proposed that a difference in the orbital shape of galaxies in FGs could justify the large magnitude gap. In fact, massive satellites found on more radial orbits merge faster with the BCG. This hypothesis got a first observational confirmation in \citet{Zarattini2021}, who measured the dependence of the anisotropy parameter on $\Delta m_{12}$. They showed that systems with $\Delta m_{12} < 1.5$ mag are characterised by isotropic orbits (e.g. galaxies are equally distributed on radial and tangential orbits), whereas systems with $\Delta m_{12} > 1.5$ mag have a larger fraction of galaxies onto radial orbits, especially in the external regions, within $0.7-1 \,R_{200}$.

The location of FGs in the large-scale structure could also play an important role to understand these systems, but few studies were devoted to this topic. \citet{Adami2012,Adami2018} studied two FGs and one non-FG with different approaches, finding FGs in a less dense environment with respect to the control group. More recently, \citet{Zarattini2022} analysed the large-scale structure around a sample of 16 FGs. These are found close to filaments, with an average distance of $(3.7 \pm 1.1)$ $R_{200}$ and a minimum distance of 0.05 $R_{200}$, and far from intersections, with an average distance of $(19.3 \pm 3.6)$ $R_{200}$ and a minimum distance of 6.1 $R_{200}$.

The goal of this paper is to analyse how these distances compare to those of non-FGs. Here, we consider a sample of 55 clusters and groups with $\Delta m_{12} < 2$ mag and we compare its properties with those of the 16 FGs analysed by \citet{Zarattini2022}. This work is part of the Fossil Group Origins project \citep[FOGO,][]{Aguerri2011}, a large observational effort aimed at characterising the properties of the sample of 34 FG candidates presented in \citet{Santos2007}. 

The cosmic web description that we use in this work comes from \citet{Chen2015,Chen2016}. The detection is based on the Subspace Constrained Mean Shift (SCMS) algorithm. The SCMS algorithm is a gradient ascent method that models filaments as density ridges, one-dimensional smooth curves that trace high-density regions within the point cloud. The algorithm consists of three steps \citep[see][for a detailed description of the formalism]{Chen2015}, the first being to estimate the underlying density function given the observed location of galaxies using the standard kernel density estimator. The second step is the denoising of the estimated density function, needed to eliminate the effect that galaxies in low-probability density regions would have on filament estimation. This step is crucial to increase the statistical power of the SCMS algorithm in low-density regions. Finally, the third and final step is the application of the original SCMS algorithm \citep{Ozertem2011} to galaxies in high-density regions. \citet{Chen2016} applied this method to a simulated dataset based on the Voronoi model \citep{vandeWeygaert1994}, showing that the SCMS algorithm reproduce clusters, filaments, and walls in a precise way. Then, the SCMS algorithm was applied in \citet{Chen2016} to SDSS DR7 and DR12 data to get the final catalogue of filaments and intersections that we use in this work. The main disadvantage of this method is that the coverage depends sensitively on the number of galaxies in the analysed sample. This could be an issue since SDSS spans a large area and the coverage can not be homogeneous in the full footprint and in the entire redshift range. For example, a well known issue of SDSS is that the selection function for the spectroscopic follow up is split in two parts \citep[see fig. 3 of][]{Tarrio2020}, with two peaks in the number of observed redshift at $z \sim 0.1$ and $z \sim 0.5$ and with a clear lack of redshift in the range $0.2 < z < 0.4$.

The cosmology adopted in this paper, as well as in the other FOGO papers, is $H_0 = 70$ km s$^{-1}$ Mpc$^{-1}$, $\Omega_{\rm m} = 0.3$, and $\Omega_\Lambda = 0.7$. We denote $R_\Delta$ as the radius of a sphere within which the average mass density is $\Delta$ times the critical density of the universe at the redshift of the galaxy system, $\theta_\Delta$ is the corresponding angular radius, while $M_\Delta$ is the mass contained in $R_\Delta$.

\section{Sample}
\label{sec:sample}

\citet{Zarattini2022} presented the analysis of the large scale environment of a sample of genuine FGs. This sample consists in the 16 FGs in the redshift range $0.03 < z < 0.15$ discussed by \citet{Aguerri2021}. However, one of these FGs, namely AWM4, was found at $z < 0.05$, which is the limit of the \citet{Chen2016} filament and intersection catalogue used in \citet{Zarattini2022} and in this work. We then remove AWM4 from the sample and remain with 15 genuine FGs.

The goal of this paper is to compare the properties of the FGs discussed in \citet{Zarattini2022} with those of a sample of non-FGs spanning similar ranges in redshift and mass. The non-FG systems are taken from \citet{Aguerri2007} and, in particular, we use all the systems with a spectrosopically confirmed magnitude gap in the redshift range $0.05 < z < 0.15$ and with $\Delta m_{12} < 2.0$ mag. The total number of these systems is 55. 

The magnitude gap distribution of FGs and non-FGs is shown in the first panel of Fig. \ref{fig:sample}. Most systems have small magnitude gaps, with a sort of a lack of systems in the range $1.5 < \Delta m_{12} < 2.0$ mag.

The mass distribution of the two samples is shown in the second panel of Fig. \ref{fig:sample}. The mass is obtained by filtering X-ray maps from the ROSAT All-Sky Survey \citep{Truemper1993,Voges1999} with the X-ray matched filter described in \citet{Tarrio2016,Tarrio2018}, which assumes the average gas density profile from \citet{Piffaretti2011}. We use the filter in fixed-position and blind-size mode, i.e. centered at the position of the object (with a small margin to search for the X-ray peak) and using 32 different sizes covering $\theta_{500}$ from 0\farcm94 to 35\farcm31. The output of the filter is a flux-size degeneracy curve, which provides the X-ray flux of the object within $R_{500}$ in the [0.1-2.4] KeV band ($F_{\rm X}$) for the different values of $\theta_{500}$. The mass is then obtained by computing the Sunyaev-Zel'dovich (SZ) flux of the cluster within $R_{500}$ ($Y_{500}$) from the $F_{\rm X}/Y_{500}$ relation found by \citet{PlanckIntI2012} at the cluster redshift, and breaking the flux-size degeneracy with the $M_{500} - D^2_A Y_{500}$\footnote{$D_A$ is the angular size distance to the galaxy system.} scaling relation from \citet{Planck2013ResXX}, which relates $\theta_{500}$ and $Y_{500}$ when $z$ is known, as explained in \citet{Planck2013ResXXIX} and \citet{Tarrio2018}. The mass distributions of FGs and non-FGs are similar to each other, although in FGs we are missing the very low-mass end of the distribution ($M_{500} < 3\times10^{13} {\rm M}_\odot$). A Kolmogorov-Smirnov (KS) test confirms that the two distributions are coming from the same parent distribution ($p_{M_{500}} = 0.99$).

From the same analysis, we also estimate the $R_{500}$ radii for both samples. This was done in order to have homogeneous measurements, since some of the radii available in the literature were computed from X-ray and others from the velocity dispersion of galaxies. We then convert the $R_{500}$ to $R_{200}$ using $R_{200} = 1.516 \times R_{500}$ \citep{Arnaud2005}. The median values of $R_{200}$ is $(1.1\pm 0.2)$ Mpc and $(1.1\pm 0.3)$ Mpc for fossils and non-fossils, respectively.

The redshift distribution of the two samples is presented in the third panel of Fig. \ref{fig:sample}. On average, FGs are found at a slightly higher redshift. However, this is not an issue since the median redshifts are $0.08\pm0.02$ and $0.12\pm0.03$ for non-FGs and FGs respectively. This corresponds to an age difference of about 0.5 Gyr, for which we do not expect any evolution effect since it is a relatively short timescale. In this case, the KS test confirms the difference between the two redshift distributions ($p_z = 0.0015$).

Finally, in the fourth panel of Fig. \ref{fig:sample} we show the distribution of the absolute magnitude of the BCGs for FGs and non-FGs. In this case, non-FGs show a clear peak at about $M_r = -23$ mag, whereas FGs show a double peak at about $M_r = -22.5$ and $M_r = -23.5$ mag. The KS test confirms the difference between the two distributions ($p_{Mr} = 0.039$).

\begin{figure}
    \centering
    \includegraphics[trim=120 420 220 80,width=0.5\textwidth]{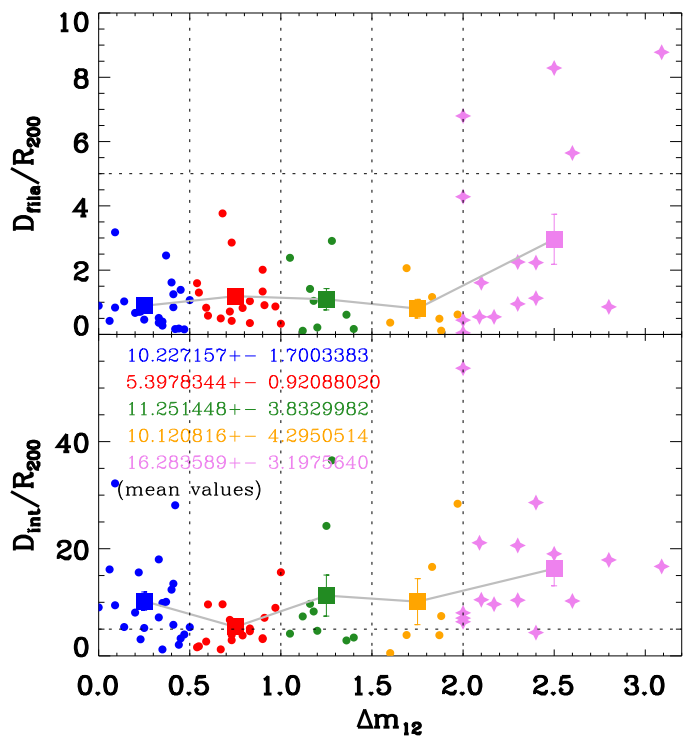}
    \caption{Minimum distance to the filaments (top panel) and intersections (bottom panel) as a function of the magnitude gap. The larger symbols correspond to the average values of the different intervals of magnitude gap as listed in Table \ref{tab:mean_values}. The horizontal dashed line in both panels indicates the distance of 4 $R_{200}$.}
    \label{fig:d12_dist}
\end{figure}

\section{Results}
\label{sec:results}
Following \citet{Zarattini2022}, we compute the distance from each system to the cosmic web filaments and intersections listed in the catalogue by \citet{Chen2016}. 
They are identified using Sloan Digital Sky Survey data in combination with the Subspace Constrain Mean Shift (SCMS) algorithm \citep[][and references therein]{Chen2015}. In particular, SCMS performs a three-step analysis (density estimation, thresholding, and gradient ascent) to detect filaments and intersections. The catalogue uses galaxies with spectroscopic redshifts in the range $0.05 < z < 0.7$.

In Fig. \ref{fig:d12_dist} we present the distances of our sample of 55 non-FGs combined with the 15 FGs analysed in \citet{Zarattini2022}  to filaments (top panel) and intersections (bottom panel) as a function of $\Delta m_{12}$. It is worth noticing that we also recompute the distances to filaments and intersections for the sample of FGs. This is due to the choice of deriving $R_{200}$ in a homogeneous way for all the analysed systems.

Non-FGs show a more compact distribution, especially when analysing the distribution of the distances to filaments, with only one of them having a filament at more than 4 $R_{200}$ ($< 2\%$). On the other hand, 5 out of 15 FGs have filaments at a larger distance than 4 $R_{200}$ (31\%). This can suggest a sort of bimodality for FGs. To test this scenario we applied a KMM test \citep{Ashman1994} to the data, finding that the bimodality is confirmed (the null hypothesis is rejected with a p-value of 0.005 and the two peaks are found at 1.1 and 6.7 $R_{200}$, respectively). However, \citet{Ashman1994} claimed that using small datasets could lead to unreliable conclusions on the bimodality and in this case the number of elements used to run the KMM test is 15.

We report the average distances from filaments as a function of $\Delta m_{12}$ in the first column of Table \ref{tab:mean_values}. 
It can be seen that, if we compare the average values for FGs and non-FGs ($0.0 \le \Delta m_{12} < 2.0$), the difference is significant at more than 2$\sigma$ level. We also run a test to compute the Student’s T-statistic and the probability that two distributions have significantly different averages ({\it t-test}). The probability is $p_{fila,ave}=0.0002$, confirming that the two distributions have different averages.

A similar result is found for the distance to intersections of the cosmic web. These distances are shown in Table \ref{tab:mean_values} too. The distance to intersections is always larger than that to filaments in all the magnitude-gap bins. This is expected, since different filaments converge to a single intersection and so it is more probable to be close to a filament than to an intersection. Moreover, \citet{Chen2016} defined filaments also within intersections, so there is always at least one filament at any intersection position.
Again, the distances to intersections for non-FGs are systematically smaller than for FGs and no FG has a distance smaller than 4 $R_{200}$ to the closest intersection. On the other hand, the smallest distances for the four non-FG bins of growing magnitude gap are 1.2, 1.2, 2.9, and 0.5 $R_{200}$, respectively.
We also run the {\it t-test} for the two distributions of distances for FGs and non-FGs, finding $p_{int,ave}=0.007$, again confirming that these distributions have different averages.
\begin{figure*}
    \centering    \includegraphics[trim=70 430 90 170,width=\textwidth]{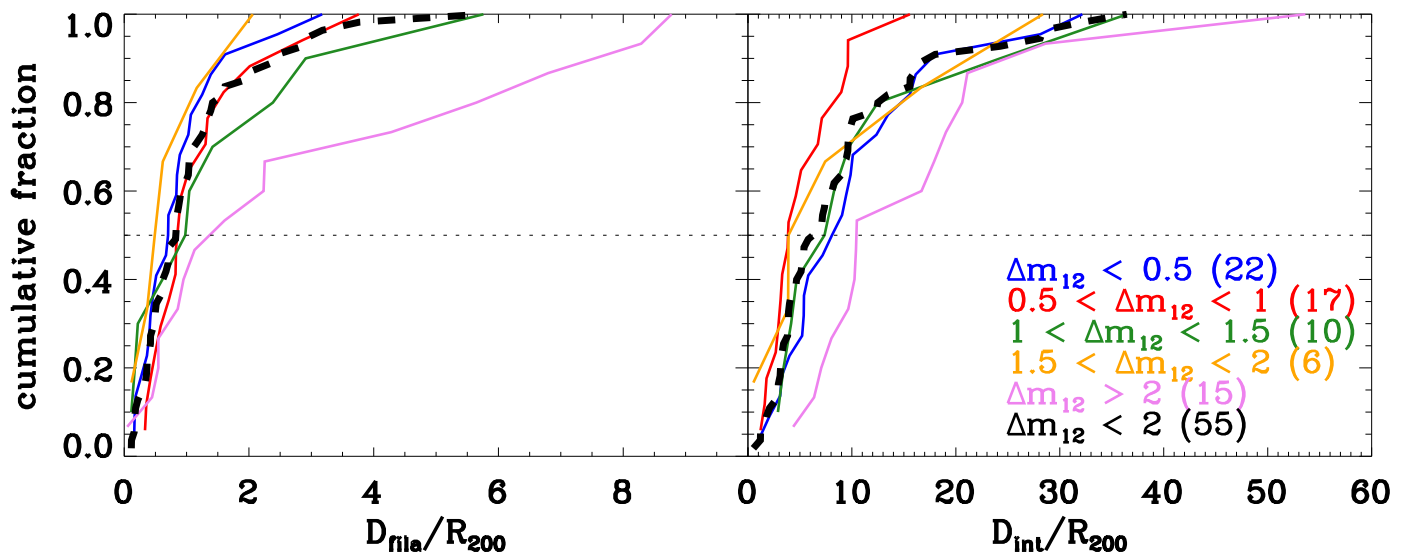}
    \caption{Cumulative distributions of distances from filaments (left panel) and intersections (right panel). Blue lines mark the systems with $\Delta m_{12} < 0.5$ mag, red lines systems with $0.5 \le \Delta m_{12} < 1.0$ mag, green lines systems with $1.0 \le \Delta m_{12} < 1.5$ mag, orange lines systems with $1.5 \le \Delta m_{12} < 2.0$ mag. In both panels, the black dashed line is the cumulative distribution for all non-FG systems ($\Delta m_{12} < 2.0)$ while violet line correspond to FGs ($\Delta m_{12} \ge 2.0$ mag). The numbers within parenthesis indicate the size of the samples for each cumulative function.} The horizontal dashed line indicates the 50\% of the distributions.
    \label{fig:cumulative}
\end{figure*}

We also compute the cumulative fraction of the distance to filaments and intersections for non-FGs in the four bins of $\Delta m_{12}$ and for FGs. The results are shown in Fig. \ref{fig:cumulative}. FGs have a different behaviour than non-FGs. We use the KS test to check if the differences are statistically significant. For filaments, the distribution of FGs is not different from that of all the non-FG bins. The result of the KS test between FGs and non-FGs is $p_{\rm fila} =0.083$. If we remove the only non-FGs that is found with a filament at more than 4$R_{200}$, the KS probability drops to $p_{\rm fila} =0.060$. This is not a statistically-significant difference, but it is very close to. This result could be related with the observed bimodality in the distance to the filaments distribution for FGs. The fact that about 2/3 of them are found close to filaments and 1/3 is at more than 4 $R_{200}$ could justify the not-significant difference in this parameter. On the other hand, the difference in the distributions of the distance to intersection is statistically significant with $p_{\rm int} = 0.013$.

\begin{table}[t]
\setlength{\tabcolsep}{15pt}
\caption{Average distances and uncertainties to filaments and intersections for different magnitude gaps.}
\label{tab:mean_values}
\begin{center}
\tiny
\begin{tabular}{lcc}
\hline 
\noalign{\smallskip}   
\multicolumn{1}{l}{Magnitude gap} & \multicolumn{1}{c}{$D_{{\rm fila}}$} & \multicolumn{1}{c}{$D_{{\rm int}}$}\\ 
\multicolumn{1}{c}{[mag]} & \multicolumn{1}{c}{[$R_{200}$]} & \multicolumn{1}{c}{[$R_{200}$]} \\
\noalign{\smallskip}  
\hline
$0.0 \le \Delta m_{12} < 0.5$ & $0.9 \pm 0.2$ & $10.2 \pm 1.7$ \\
$0.5 \le \Delta m_{12} < 1.0$ & $1.2 \pm 0.2$ & $5.4 \pm 0.9$ \\
$1.0 \le \Delta m_{12} < 1.5$ & $1.6 \pm 0.6$ & $11.4 \pm 3.4$ \\
$1.5 \le \Delta m_{12} < 2.0$ & $0.8 \pm 0.3$ & $10.1 \pm 4.3$	\\
$0.0 \le \Delta m_{12} < 2.0$ & $1.1 \pm 0.1$ & $8.9 \pm 1.1$ \\
$\Delta m_{12} \ge 2.0$ & $3.0 \pm 0.8$ & $16.3 \pm 3.2$ \\
\hline
\end{tabular}
\end{center}   
\tablefoot{ Column (1): Magnitude gap. Column (2): Arithmetic average distance to filament in units of $R_{200}$. Column (3): Arithmetic average distance to intersection in units of $R_{200}$. The reported uncertainties in columns (2) and (3) are the standard errors of the averages, which we calculated as the standard deviation of the measured values divided by the square root of the number of measurements.}
\end{table} 

\section{Discussion and conclusions}
\label{sec:discussion}
In Fig. \ref{fig:sample} we show the mass distribution of FGs and non-FGs. The distributions are quite similar, but we also compute the cumulative distribution of distances removing the most-extreme systems in both samples. In particular, we repeated the computation using only objects in the mass range $10^{13} \le M_{500} \le 2.5\times 10^{14}$ M$_\odot$. The results remains unchanged, being the distance distribution from intersections statistically different between FGs and non-FGs ($p_{\rm int} = 0.026$). No differences are found for filaments ($p_{\rm fila} = 0.14$).

Our result can be interpreted in terms of the formation scenario of FGs. In fact, \citet{Ponman1994} claimed that FGs were able to build large magnitude gaps thanks to their isolation from the cosmic web. Indeed, in this work we show that FGs are systematically more isolated than non-FGs. In particular, the average distance to filaments and intersections is larger for FGs than non-FGs. The distance to intersections is of particular interest, since intersections are considered as the places in the cosmic web where the mass accretion is more efficient and where galaxy clusters are supposed to live. However, we are now showing that this distance is statistically different between FGs and non-FGs, being FGs systematically farther to intersections than non-FGs.

\citet{Kraljic2018} showed that filament galaxies at more than 3.5 Mpc from the centre of a intersection are not suffering (or marginally suffering) the influence of the intersection itself. This measure is referred to isolated galaxies and not to groups and clusters. However, we speculate that the larger distance that separates FGs from intersections can have an impact on their mass-accretion process (e.g. a smaller number of galaxies could be accreted far from intersections).
It is important to keep in mind that we are computing the distances from the centres of filaments and intersections, for which \citet{Chen2016} did not estimate the size. Thus, it is difficult to say, with the available data, if our FGs are really outside filaments and intersections. We can only claim that FGs are systematically farther than non-FGs in both cases.

In \citet{Cossairt2022} the large-scale structure around FG- and non-FG-candidates was studied using a gravity-only cosmological simulation. The authors found that there is no statistical difference in the environment where FGs and non-FGs formed. However, it is worth noticing that the selection of FGs in \citet{Cossairt2022} is deeply different from ours. In particular, we use the observational definition based on the magnitude gap, whereas they did not use this parameter and their selection is based on the idea that FGs are old systems that have not suffered major mergers in recent times. However, this idea does not properly describe the optically-selected FGs. In fact, it was shown that systems selected from the magnitude gap are not particularly old and their BCGs have similar stellar populations as non-FGs \citep{LaBarbera2009,Eigenthaler2013,Corsini2018,Chu2023}, their fraction of substructure is similar to that of non-FGs \citep{Zarattini2016}, and the building of the gap is very recent \citep[][]{Kundert2017}. For optically-selected FGs \citet{Jones2003} and \citet{Adami2018,Adami2020} found hints of a different large-scale environment, being FGs more isolated (e.g. in less dense environment): a result that we are now confirming with a statistically-significant sample. We thus think that a direct comparison between optically-selected FGs and those in the simulations by \citet{Cossairt2022} is not straightforward due to the very different selection functions. Indeed, the latter systems are probably more connected to the original idea of what an FG was expected to be. Nevertheless, we can learn a lot from this difference. In particular, we can now say that the comparison between observational findings and simulation results asks for changing the definition of ``fossil groups'' itself: systems selected with a large magnitude gap at $z=0$ are not old fossils of the ancient Universe, but they are isolated systems 
 with a peculiar position in the cosmic web. This location reduces the number of major accretion events in recent epochs and makes the formation of the magnitude gap to be connected to the internal evolution of these systems, with their high merging rate supported by more radial orbits \citep{Zarattini2021}. On the other hand, \citet{Cossairt2022} showed that systems selected to be old and relaxed are not in a peculiar location in the cosmic web and they have to be selected with a different diagnostic than the magnitude gap.
Indeed, \citet{Kundert2017} studied a sample of FGs selected in the Illustris cosmological simulations for having $\Delta m_{12} > 2.0$ mag at $z=0$ and they compared the halo mass assembly at early times for these FGs with a sample of non-FGs, finding no differences in their formation times. This result confirmed that the magnitude gap is not the proper parameter to select old systems. 

In a very recent work, \citet{Taverna2023} have analysed the large-scale structure around a sample of Hickson-like compact groups. These systems were claimed to be the precursors of FGs \citep{Barnes1989,Vikhlinin1999,Jones2003}, it is thus interesting to understand if the position of FGs and compact groups in the cosmic web is similar or not. \citet{Taverna2023} define four different environments: (i) nodes of filaments, (ii) loose groups, (iii) filaments, and (iv) cosmic voids. They claimed that 45\% of compact groups do not reside in any of these structures. This result seems in good agreement with our findings, although we already mentioned that we are not able to define if our systems are {\it inside} a filament or an intersection. However, we do find that FGs are systematically farther from both type of structures. So, according to these results, a link between FGs and compact groups can not be excluded \citep[but see also][]{MendesDeOliveira2007}.

Our conclusions can be summarised as follows:

\begin{itemize}
    \item We compared a sample of FGs taken from \citet{Aguerri2021} with a sample of 55 non-FGs taken from \citet{Aguerri2007}. The two samples have similar mass and slightly different redshift ranges. However, the latter is not significant in terms of cluster evolution, corresponding to about 0.5 Gyr.
    \item FGs are more isolated from the cosmic web than non-FGs. In fact, their average distances to filaments and intersections are $(3.0\pm0.8)$ $R_{200}$ and $(16.3\pm3.2)$ $R_{200}$, respectively. For comparison, the average distances for non-FGs are $(1.1\pm0.1)$ $R_{200}$ and $(8.9\pm1.1)$ $R_{200}$, respectively. These results remain qualitatively unchanged if we remove the most-extreme systems in terms of mass. A t-test confirmed that the two distributions have different averages.
    \item The cumulative distribution of the distances to intersections is also found to be different between FGs and non-FGs, with the former found at larger distance. The statistical significance of the result was confirmed with a KS test.
    \item For non-FGs, the distance to filaments and intersections seems not to depend on the magnitude gap. The difference arises only for systems with $\Delta m_{12} \ge 2$ mag.
    \item We can thus conclude that FGs are in a peculiar position of the cosmic web, being more isolated than non-FGs. This result is in agreement with previous observational works \citep{Jones2003,Adami2018,Adami2020}, but not with the gravity-only simulations by \citet{Cossairt2022}. The most-probable reason for this disagreement is the different selection function of FGs in observations (based on the magnitude gap at $z=0$) and simulations (lacking of massive halo mergers since $z=1$).
\end{itemize}

Our interpretation of this tension between observations and simulations is that FGs selected with $\Delta m_{12} > 2$ mag at $z=0$ are {\it not old fossils} of the ancient Universe, but {\it systems isolated} from the cosmic web. The formation of the large magnitude gap would then be related more to internal processes \citep[e.g. the prenominance of radial orbits,][]{Zarattini2021} rather than to an early formation epoch.

\section{Acknowledgement}
We thanks the referee for his/her constructive comments that improved the clarity and the quality of the paper. We also would like to thank Marisa Girardi for useful discussions about the KMM test. SZ is supported by the Ministry of Science and Innovation of Spain, projects PID2019-107408GB-C43 (ESTALLIDOS) and PID2020-119342GB-I00, by the Government of the Canary Islands through EU FEDER funding project PID2021010077. SZ and EMC are supported by MIUR grant PRIN 2017 20173ML3WW-001 and Padua University grants DOR2019-2021. JALA is supported by the grant PID2020-119342GB-I00.

\bibliography{bibliografia}

\end{document}